\documentclass[english,aip,jcp,amssymb, amsmath,reprint,superscriptaddress,showpacs]{revtex4-1}
\usepackage[T1]{fontenc}
\usepackage[utf8]{inputenc}
\usepackage{color}
\usepackage{babel}
\usepackage{array}
\usepackage{textcomp}
\usepackage{multirow}
\usepackage{amstext}
\usepackage{graphicx}
\usepackage[unicode=true,
 bookmarks=true,bookmarksnumbered=false,bookmarksopen=false,
 breaklinks=true,pdfborder={0 0 0},pdfborderstyle={},backref=false,colorlinks=true,pdfpagemode=FullScreen]
 {hyperref}
\hypersetup{pdftitle={Si(331)-(12×1) surface revisited},
 pdfauthor={R. Zhachuk, J. Coutinho, K. Palotás},
 pdfsubject={68.35.bg, 68.35.Gy, 68.35.Md},
 pdfkeywords={Silicon, Surface reconstruction},
 allcolors=blue}

\makeatletter

\providecommand{\tabularnewline}{\\}

\makeatother

\begin{document}
\title{Atomic and electronic structure of the Si(331)-(12\texttimes 1) surface}
\author{Ruslan Zhachuk}
\email{zhachuk@gmail.com}

\affiliation{Institute of Semiconductor Physics, pr. Lavrentyeva 13, Novosibirsk
630090, Russia}
\author{José Coutinho}
\affiliation{Department of Physics \& I3N, University of Aveiro, Campus Santiago,
3810-193 Aveiro, Portugal}
\author{Krisztián Palotás}
\affiliation{Institute of Physics, Slovak Academy of Sciences, Dúbravská cesta
9, 84511 Bratislava, Slovakia}
\affiliation{Wigner Research Center for Physics, Hungarian Academy of Sciences,
P.O. Box 49, 1525 Budapest, Hungary}
\affiliation{MTA-SZTE Reaction Kinetics and Surface Chemistry Research Group, University
of Szeged, Rerrich B. tér 1, 6720 Szeged, Hungary}
\date{\today}
\begin{abstract}
We report on the investigation of the atomic and electronic structure
of clean Si$(331)\textrm{-}(12\times1)$ surface using a first-principles
approach with both plane wave and strictly localized basis sets. Starting
from the surface structure proposed by Zhachuk and Teys {[}Phys. Rev.
B \textbf{95}, 041412(R) (2017){]}, we develop significant improvements
to the atomic model and localized basis set which are critical for
the correct description of the observed bias dependence of scanning
tunneling microscopy (STM) images. The size mismatch between the Si
pentamers from the surface model and those seen by STM is explained
within the context of the Tersoff-Hamann model. The energy barriers
that separate different Si$(331)$ buckled configurations were estimated,
showing that the surface structure is prone to dynamic buckling at
room temperature. It is found that empty electronic states on Si(331)
are essentially localized on the pentamers with interstitials and
under-coordinated Si \emph{sp}$^{2}$-like atoms between them, while
filled electronic states are localized on under-coordinated Si \emph{sp}$^{3}$-like
atoms and dimers on trenches. The calculated electronic density of
states exhibits two broad peaks in the fundamental band gap of Si:
one near the valence band top, and the other near the conduction band
bottom. The resulting surface band gap of 0.58~eV is in an excellent
agreement with spectroscopy studies. 
\emph{Published in Journal of Chemical Physics}\textbf{\emph{149}}\emph{,
204702 (2018).} \texttt{\textbf{\textcolor{blue}{https://doi.org/10.1063/1.5048064}}}
\end{abstract}
\maketitle

\section{Introduction}

Semiconductor surfaces often reconstruct in order to lower the surface
energy by reducing the density of dangling bonds. Some surface reconstruction
elements exhibit buckling bi-stability, eventually enabling the control
of the buckled state through the application of external fields. For
instance, it was demonstrated that the reconstructed $(100)$ surfaces
of Si and Ge contain tilted atomic-dimers whose buckled state can
be locally manipulated by means of a scanning tunneling microscope
(STM) tip.\citep{sag03,tak04} This effect led to the aspirations
of realizing atomically dense storage devices compatible with the
group-IV semiconductor technology. Despite the prospects, the controlled
switching of the buckled state was only possible at a low temperature,
which severely undermines its applicability. The main limitation lies
in the low energy barriers that separate different buckled configurations,
and, therefore, in memory loss due to the thermal motion of atoms.

Buckled structural elements were also recently found on high-index
Si$(331)$ and Si$(7\,7\,10)$ surfaces.\citep{zha17,zha14} While
the main interest in the Si$(7\,7\,10)$ surface stems from its widespread
use as a template for the growth of highly ordered nano-object arrays,\citep{zha04,tey06}
the reason for the interest in the Si$(331)$ surface is related to
the problem of phase transitions in two-dimensional systems, and that
is a fundamental problem. 

The Si$(331)$ surface is the only stable surface between Si$(111)$
and Si$(110)$.\citep{olsh81} The surface exhibits a complex reconstruction
which can only be correctly described by a matrix notation.\citep{bat09}
In practice, however, it is often referred to as $(12\times1)$ and
sometimes as $(6\times2)$,\citep{bat09,tey17} which does not correspond
to Wood's notation.\citep{woo64} The $(12\times1)$ surface reconstruction
is stable up to about $810\,\mathrm{\text{°}C}$.\citep{hib96} It
was shown that the $(12\times1)$-to-$(1\times1)$ phase transition
in this two-dimensional system is very unique since it proceeds in
two stages. At the first (lower temperature) stage, the domain size
of the reconstruction decreases as the temperature increases. This
stage is the process of domain boundary proliferation and is a continuous
phase transition. At the second (higher temperature) stage, the reconstructed
Si$(331)$ facets disappear according to a first-order phase transition.
The nature of the first stage observed during the $(12\times1)$-to-$(1\times1)$
phase transition is unclear and we expect that the explanation may
arise from the atomistic model of the Si$(331)\textrm{-}(12\times1)$
surface reconstruction. 

There were several attempts to build the atomic model for the Si$(331)\textrm{-}(12\times1)$
surface in the past. While the first models were based on simple adatoms
and dimers,\citep{olsh98,gai01} more recent versions include complex
structural elements, such as pentamers with interstitials.\citep{bat09,zha17}
It was shown recently that the 8-pentagon (8P) atomic model for the
Si$(331)$ reconstruction proposed by Zhachuk and Teys\citep{zha17}
shows a remarkably low surface energy and closely reproduces the experimental
STM images. It is interesting to note that the 8P model is the only
model of the Si$(331)$ surface which does not include adatoms. According
to the 8P model, the surface contains buckled structures,\citep{zha17}
although it was not clear how stable these buckled configurations
could be at room temperature.

In Ref.~\onlinecite{tey17} a critical remark was made against a
predecessor structure of 8P, namely that the pentamer-like features
shown in the experimental STM images are about 1.5-1.8 times larger
than the 5-fold rings of Si atoms from the atomistic model.\citep{bat09}
Since pentamers with interstitials are inherent parts of the 8P model
as well, this raises doubts regarding its correctness. Below, we demonstrate
that the above discrepancies are only apparent and are easily circumvented
if we consider comparable surface-to-tip distances for both experiments
and theory. We also present additional and compelling arguments supporting
the 8P model for the $(12\times1)$ reconstruction, namely: (i) an
excellent account for the observed bias-dependent STM images and (ii)
a very good agreement between the calculated electronic density of
states (DOS) and spectroscopic data.

The aim of the present work is to further develop the 8P model of
the Si$(331)\textrm{-}(12\times1)$ surface proposed in Ref.~\onlinecite{zha17},
reconcile the available experimental data with the Si$(331)$ structural
model and estimate the energy barriers that separate different Si$(331)$
buckled configurations. 

The paper is organized as follows: after reviewing the calculational
procedure we report on a revised 8P model of the Si$(331)$ surface
and estimate the energy barriers separating different buckled configurations.
We then explain the apparent mismatch between Si pentamers from the
surface model and those seen by STM, describe several improvements
to the local atomic basis set and its impact to the quality of the
calculated STM images. In the last section, we compare the bias-dependent
experimental and calculated STM images, DOS spectra and localization
of the states edging the Si$(331)$ surface band gap.

\section{Computational details}

The first-principles calculations reported in this work were carried
out within standard and hybrid density functional theory (DFT) using
the \textsc{siesta} and \textsc{vasp} simulation packages, respectively.\citep{sol02,kre93,kre94,kre96,kre96a}
All calculations reported in this work but the global and local density
of states (LDOS) were performed using the \textsc{siesta} software.
This code employs a local basis set, allowing us to efficiently explore
more than hundred atomic configurations with moderate computational
resources. Here we used the local density approximation (LDA) to the
exchange-correlation energy,\citep{per92} along with the norm-conserving
pseudopotentials of Troullier and Martins.\citep{tro91} The Kohn–Sham
states were described with the help of linear combinations of the
Sankey–Niklewski type atomic orbitals, which included multiple zeta
orbitals and polarization functions.\citep{sol02} The atoms from
the four upper Si bilayers of the slab contributed with two sets of
\emph{s-} and \emph{p-}orbitals plus one\emph{ d-}orbital (double-$\zeta$
polarized basis set, DZP). On the other hand, the Si atoms from the
four bottom-most bilayers of the slab had only one set of \emph{s-
}and \emph{p-}orbitals, and, finally, all passivating H atoms were
assigned a single \emph{s-}orbital (single-$\zeta$ basis set, SZ).
Such choice for the basis was previously shown to result in surface
energies at a comparable accuracy to those using a full DZP basis,
while it requires less computational resources.\citep{zha13} The
electron density and potential terms were calculated on a real space
grid with a spacing equivalent to the plane-wave cut-off of $200\,\mathrm{Ry}$.

We used 8 bilayers thick Si$(331)$ slabs terminated by hydrogen from
one side. In the bulk case, our calculation yields a cubic lattice
constant of $a_{0}=5.420$~Å. The Si atomic positions in the opposite
side were set up according to the recently proposed 8P model\citep{zha17}
of the Si$(331)\textrm{-}(12\times1)$ surface revised in Section~\ref{subsec:revising-8P}
of the results. A $10\,\mathrm{\mathring{A}}$ thick vacuum layer
was used. The rectangular surface unit cell, as outlined in Fig.~\ref{fig1}(c),
was employed. The Brillouin zone was sampled using a $4\times4\times1$
$\mathbf{k}$-point grid.\citep{mon76} The positions of all slab
atoms (except for the Si atoms in two bilayers at the bottom and all
H atoms) were fully optimized until the atomic forces became less
than $1\,\mathrm{meV/\mathring{A}}$. The constant-current STM images
were produced within the Tersoff-Hamann approach.\citep{ter85} The
\textsc{WSXM} software was used to process the calculated STM images.\citep{hor07}

Furthermore, we performed DOS and LDOS calculations using the \textsc{vasp}
simulation package and hybrid density functional of Heyd-Scuseria-Ernzerhof
(HSE06).\citep{hse06,kru2006} This allows us to circumvent the well-known
insufficiencies of local and semi-local exchange-correlation functionals
in the description of the band structure and directly compare our
data with angle-resolved photoelectron spectroscopy and scanning tunneling
spectroscopy measurements.\citep{bat11} The HSE06 functional mixes
the semi-local ($E_{\mathrm{x}}^{\mathrm{PBE}}$) and Hartree-Fock-like
exact exchange $E_{\mathrm{x}}^{\mathrm{HF}}$ interactions at short
ranges (SR), treating the long-range (LR) exchange within the simpler
generalized gradient approximation (GGA) as proposed by Perdew, Burke,
and Ernzerhof (PBE),\citep{per96}

\begin{equation}
E_{\mathrm{x}}^{\mathrm{HSE06}}=\frac{1}{4}E_{\mathrm{x}}^{\mathrm{HF,SR}}(\mu)+\frac{3}{4}E_{\mathrm{x}}^{\mathrm{PBE,SR}}(\mu)+E_{\mathrm{x}}^{\mathrm{PBE,LR}}(\mu),
\end{equation}
where the separation between SR and LR parts is set at $1/\mu=5$~Å.
The correlation is fully accounted for within the PBE level. Accordingly,
the calculated indirect band gap of silicon (as obtained from the
Kohn-Sham data), using the HSE06 functional, was $E_{\mathrm{g}}=1.20$~eV.
This is major improvement over $E_{\mathrm{g}}=0.55$~eV from a PBE
calculation, and accounts very well for $E_{\mathrm{g}}=1.17$~eV
from the measurements at liquid helium temperature. The calculated
lattice parameter of Si within HSE06 was $a_{0}=5.435$~Å. In these
calculations, the core electrons are replaced by projector augmented
wave (PAW) potentials,\citep{blo94,kre99} while the valence was described
using plane waves with the kinetic energy up to $E_{\mathrm{cut}}=250$~eV.
The relaxation of atomic coordinates was performed within the PBE
level until the maximum force acting on atoms was below the threshold
of 5~meV/Å. Slab geometries, $\mathbf{k}$-point sampling and the
resulting surface structures were essentially the same as those obtained
using the local basis code. A few test calculations comparing the
STM images and energy barriers from \textsc{vasp}-PBE and \textsc{siesta}-LDA
are reported below.

\section{Results and discussion}

\subsection{Revising the 8P model of the Si$(331)\textrm{-}(12\times1)$ surface\label{subsec:revising-8P}}

\begin{figure}
\includegraphics[width=6.5cm]{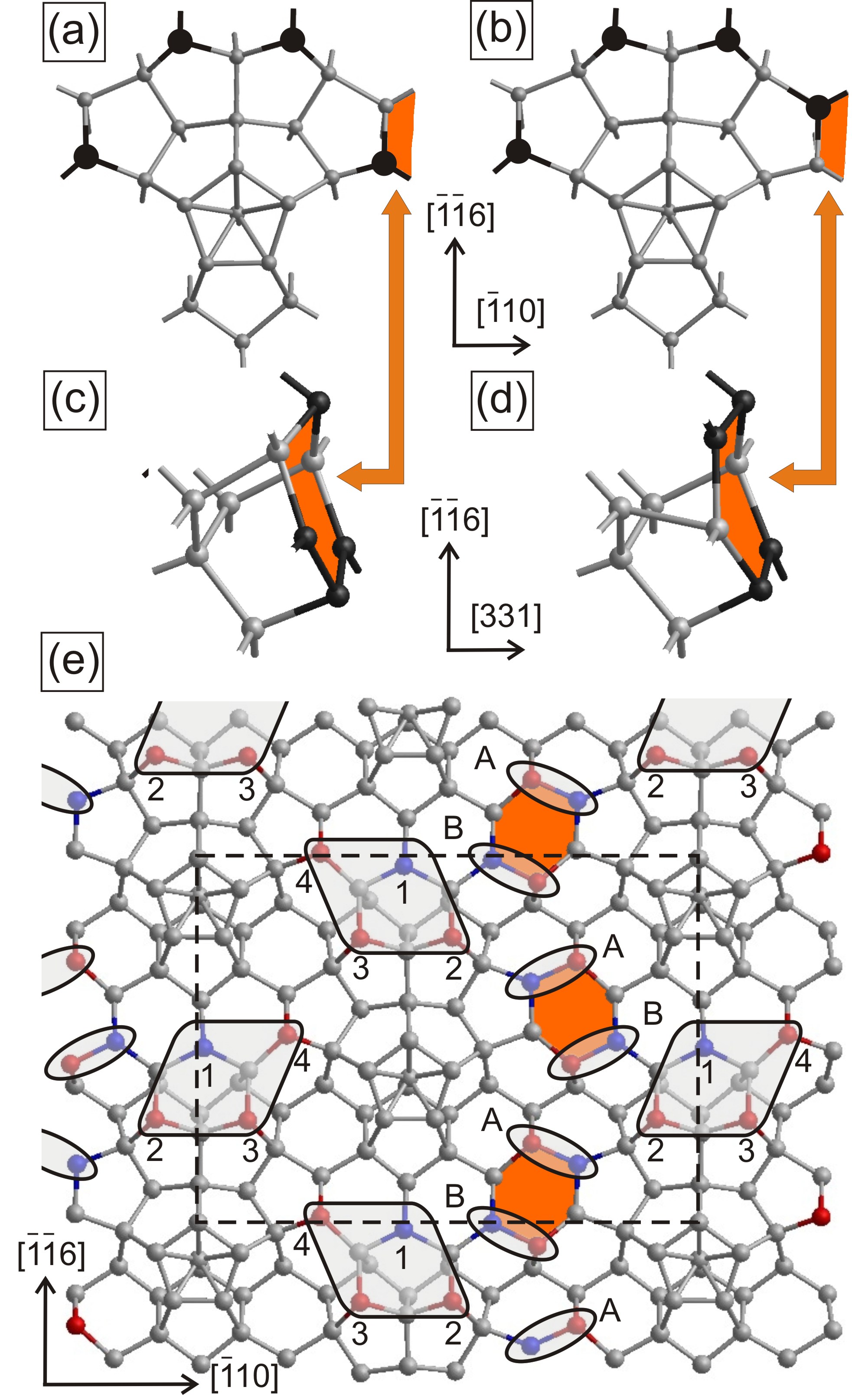}

\caption{\label{fig1}(a) and (b): Elementary building block structure of the
Si$(331)\textrm{-}(12\times1)$ surface: eight-pentagon unit (8PU).
(c) and (d): Perspective view of the area, adjacent to 8PU, demonstrating
rebonding between surface atoms and a Si atom in the deeper layer
(see the text). The atoms with dangling bonds are shown as black-filled
circles. (a) Original 8PU structure with the symmetric arrangement
of under-coordinated Si atoms (Ref.~\onlinecite{zha17}) and (c)
its respective adjacent area. (b) Revised 8PU structure with the broken
mirror symmetry proposed in the present work and (d) its respective
adjacent area. (e) Revised 8P model of the Si$(331)\textrm{-}(12\times1)$
surface. The lowest energy buckled configuration is shown. The $(12\times1)$
unit cell is outlined by a dashed line. Red/blue balls indicate raised/lowered
under-coordinated Si atoms. The parts of the structure which are prone
to buckling include four atom units (FAU) and dimers highlighted by
rhombs and ovals, respectively. The atoms in FAU are numbered 1-4,
two types of dimers are marked by ``A'' and ``B''. Orange-shaded
areas in (a)-(e) mark identical surface areas and are depicted for
eye guidance.}
\end{figure}
The structure of the 8-pentagon unit (8PU), proposed as the main structural
building block of the Si$(331)\textrm{-}(12\times1)$ surface,\citep{zha17}
is shown in Fig.~\ref{fig1}. The structure has mirror symmetry in
the vertical plane ($[\bar{1}\bar{1}6]$ direction). This is also
valid for the arrangement of under-coordinated Si atoms in 8PU (black-filled
circles in Fig.~\ref{fig1}(a)). The revised version of 8PU is shown
in Fig.~\ref{fig1}(b). The difference between the original and revised
versions is that the arrangement of under-coordinated Si atoms within
the 8PU in Fig.~\ref{fig1}(b) is not symmetric. The transformation
from the atomic configuration shown in Fig.~\ref{fig1}(a) into that
shown in Fig.~\ref{fig1}(b) consists of the simultaneous bond breaking
(formation) between the upper (lower) atom in the right-hand side
of 8PU and the Si atom in the deeper layer (see Figs.~\ref{fig1}(c)
and \ref{fig1}(d)). According to the revised 8P model of the Si$(331)$
surface, the rebonded sides of 8PUs are facing the trenches between
the pentamers which are observed as dark vertical stripes in experimental
STM images (Figs.~\ref{fig2}(a) and \ref{fig2}(b)).

The location of the revised 8PUs on the reconstructed Si$(331)$ surface
is shown in Fig.~\ref{fig1}(e). Due to the rebonded Si atoms in
8PU described above, the trench between 8PUs in the revised surface
model contains Si buckled dimers only. They are highlighted by ovals
in Fig.~\ref{fig1}(e). There are three main reasons for revising
the 8PU structure. First, the surface energy of the Si$(331)\textrm{-}(12\times1)$
with the revised 8PUs, shown in Fig.~\ref{fig1}(c), is $1.2\,\mathrm{meV/\mathring{A}^{2}}$
($38.4\,\mathrm{meV/(1\times1)\,cell}$) lower than that of the original
8PUs reported in Ref.~\onlinecite{zha17}. Second, the original 8PUs
fail to reproduce the weak spots in the STM images of the trenches,
marked by the white down triangle in Fig.~\ref{fig2}(b) (see the
Appendix for the calculated STM images according to the original 8P
model). These spots are clearly seen in the experimental STM images
available in the literature.\citep{bat09,tey17} Such a limitation
in the original 8PU model stems from the Si atoms on the positions,
which being fully coordinated, become virtually invisible in STM.
This effect is accounted for by the revised 8PU. The third reason
is an excellent agreement between the spectroscopy data\citep{bat11}
and the calculated surface band gap for Si$(331)$ when using the
revised 8P model, as shown is Section~\ref{subsec:contrasts}. Conversely,
the original 8P model proposed in Ref.~\onlinecite{zha17} shows
a non-zero DOS in the middle of the band gap (see the Appendix section
for the calculated DOS according to the original 8P model).

As already noted in Ref.~\onlinecite{zha17}, the 3-fold coordinated
atoms of the Si(331) surface buckle by shifting either out of the
surface or into it. The lowest energy buckled configuration of the
Si$(331)\textrm{-}(12\times1)$ surface with the revised 8PUs is shown
in Fig.~\ref{fig1}(e). The protruded and depressed atoms are marked
by red and blue balls, respectively. The surface unit cell contains
two sets of identical reconstruction elements (8PUs with surroundings).
The structural parts, which are prone to buckling, include four atom
units (FAUs) and two types of dimers highlighted by rhombs and ovals
in Fig.~\ref{fig1}(e).

\begin{table}
\begin{ruledtabular}
\caption{\label{tab1}Metastable buckled configurations, relative surface energies
$E_{\mathrm{surf}}$ in $\mathrm{meV/\mathring{A}}^{2}$ and in $\mathrm{meV/(1\times1)\,cell}$
(in brackets), calculated LDA band gaps $E_{\mathrm{gap}}^{\mathrm{LDA}}\,\mathrm{(eV)}$
and energy barriers $E_{\mathrm{a}}\,\mathrm{(eV)}$, separating initial
$M_{\mathrm{i}}$ and final $M_{\mathrm{f}}$ buckled configurations
of the Si$(331)\textrm{-}(12\times1)$ surface. Buckled configurations
have labels of type `xxxx-xx', where the first four letters describe
the state of atoms in FAU, while the last two letters show the state
of dimers in the trench. `x' can be either ''n'' or ``f'', where
``n'' means that the structural element is in its ``normal'' state,
corresponding the lowest energy configuration shown in Fig.~\ref{fig1}(c),
while ``f'' means that the element is in the opposite state, ``flipped''.}
\begin{tabular}{cccccc}
$M_{\mathrm{i}}$ & Configuration & $E_{\mathrm{surf}}$ &  $E_{\mathrm{gap}}^{\mathrm{LDA}}$ & $M_{\mathrm{f}}$ &  $E_{\mathrm{a}}$\tabularnewline
\multirow{2}{*}{1} & \multirow{2}{*}{nnnn-nn} & \multirow{2}{*}{0 (0)} & \multirow{2}{*}{0.33} & 3 & 0.29\tabularnewline
 &  &  &  & 4 & 0.21\tabularnewline
2 & nnnn-ff & 1.8 (57.6) & 0.23 & 3 & 0.01\tabularnewline
\multirow{2}{*}{3} & \multirow{2}{*}{nnnn-nf} & \multirow{2}{*}{1.4 (44.8)} & \multirow{2}{*}{0.26} & 1 & 0.03\tabularnewline
 &  &  &  & 2 & 0.09\tabularnewline
4 & fnfn-nn & 0.8 (25.6) & 0.17 & 1 & 0.06\tabularnewline
5 & ffnf-fn & 2.2 (70.5) & 0.07 & \multicolumn{2}{c}{No data}\tabularnewline
6 & ffnf-ff & 2.8 (89.7) & 0.10 & \multicolumn{2}{c}{No data}\tabularnewline
7 & ffnf-nf & 3.1 (99.3) & 0.11 & \multicolumn{2}{c}{No data}\tabularnewline
\end{tabular}
\end{ruledtabular}

\end{table}
We have studied the relative surface energies of different buckled
configurations of Si$(331)$ and the energy barriers separating them.
The results are given in Table~\ref{tab1}. The energy barriers reported
in Tab.~\ref{tab1} assume the transformation of one of the two 8PUs
with the surroundings, while the second 8PU in the unit cell remains
unchanged. The energy barriers are calculated by constructing 5 intermediate
stages between two chosen buckled states and linear interpolating
all atomic coordinates between them. The number of intermediate structures
was large enough, and a smooth energy profile across neighboring images
was obtained. This procedure guarantees that the value of the real
energy barrier is below the calculated one. Moreover, test calculations
show that the values of the energy barriers obtained using the linear
interpolation of atomic coordinates are very close to that calculated
using the nudged elastic band method.\citep{hen00} This is due to
the short distances between end-structures in the configurational
space. For the gentle atomic transformation like buckling considered
in this work, the difference between the energy barrier values calculated
using the two methods does not exceed $0.03\,\mathrm{eV}$. 

We have explored \emph{all} possible buckled configurations within
the same topology of covalent bonds, as shown in Fig.~\ref{fig1}(c).
The stable buckled configurations found in this study are listed in
Table~\ref{tab1}. Accordingly, the calculated energy barriers separating
different buckled configurations are all below $0.3\,\mathrm{eV}$. 

It is possible to estimate how high the kinetic barriers have to be
in order make buckled configurations stable at specific operating
temperatures. To that end we make use of the frequency $\left(f\right)$
of thermally activated switching of the buckled state using an Arrhenius
relation $f=f_{0}\exp\left(-E_{a}/kT\right)$, where $k$, $T$ and
$f_{0}$ stand for the Boltzmann constant, sample temperature and
attempt frequency, which can be approximated by the Debye frequency
of Si $(14\,\mathrm{THz})$. Assuming stable structures as those flipping
slower than $1\,\mathrm{Hz}$, we define a threshold energy barrier,
$E_{\mathrm{a,th}}=kT\ln(f_{0})$, below which the structure flips
at a rate faster than 1~Hz. Hence, at room temperature $(300\,\mathrm{{^\circ}K})$,
liquid nitrogen temperature $(77\,\mathrm{{^\circ}K})$ and liquid
helium temperature $(4\,\mathrm{{^\circ}K})$, the corresponding threshold
energy barriers are $E_{\mathrm{a,th}}=0.78\,\mathrm{eV}$, $0.20\,\mathrm{eV}$
and $0.01\,\mathrm{eV}$.

Considering that $E_{\mathrm{a}}\leq0.3$~eV, similarly to the behavior
of dimers on the Si$(100)$ surface, we expect that both FAU atoms
and dimers on Si(331) should buckle dynamically at room temperature
due to the thermally activated motion of atoms.\citep{ham86} On the
other hand, the energy barrier separating the original 8PU from the
revised 8PU shown in Figs.~\ref{fig1}(a) and (b), respectively,
which involves the breaking and formation of covalent bonds, is significantly
higher ($\approx0.8\,\mathrm{eV}$). Therefore, we may expect that
a spontaneous rebonding on the Si$(331)$ surface at room temperature
is less likely. 

As shown in Tab.~\ref{tab1}, configurations with lower surface energy
show wider band gap. This well agrees with the general conclusions
made for semiconductor surfaces.\citep{bech03}

\subsection{Atomic orbitals for the calculation of bias-dependent STM images}

\begin{figure}
\includegraphics[width=6.5cm]{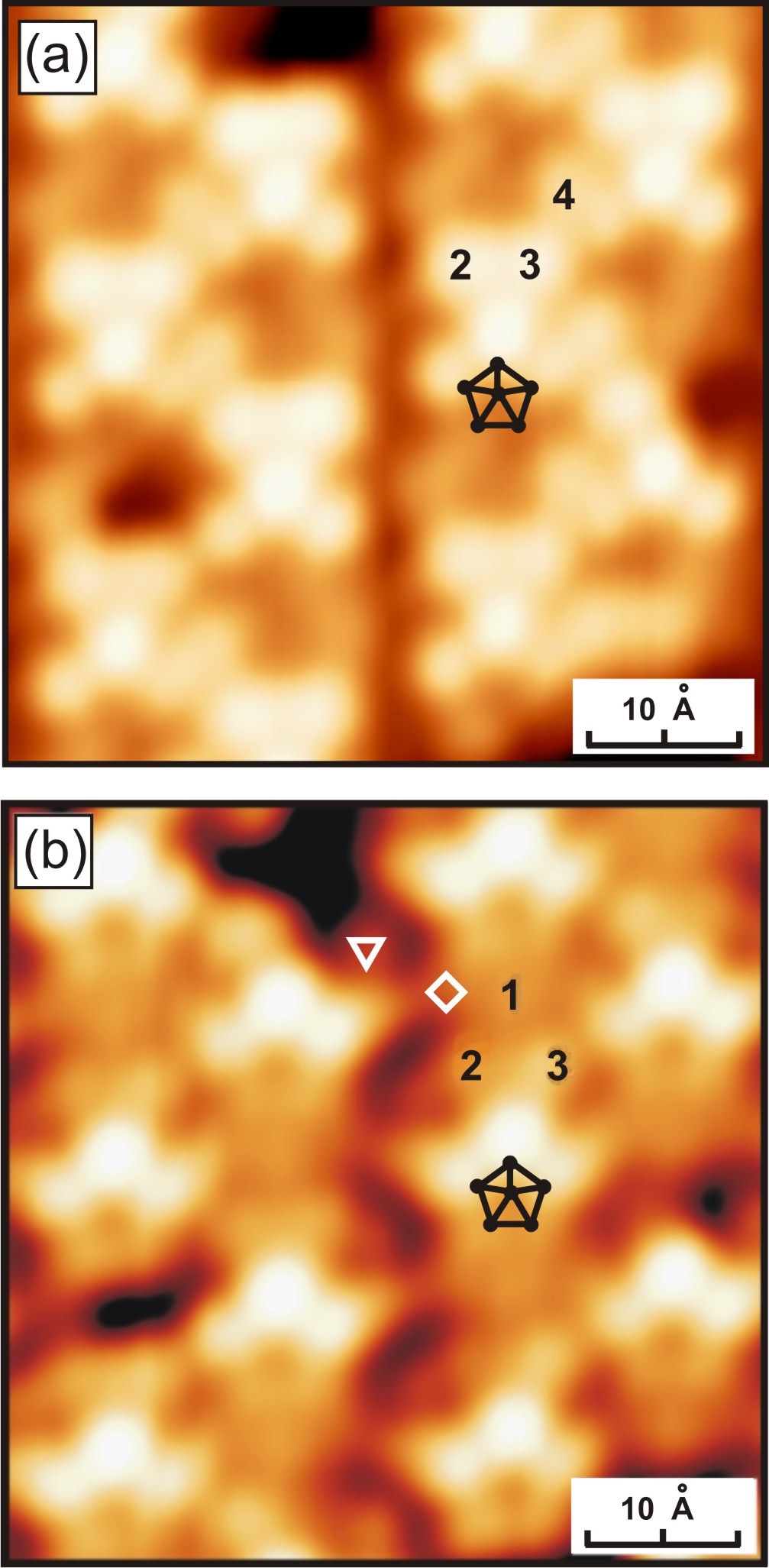}

\caption{\label{fig2} Experimental STM images of the Si$(331)\textrm{-}(12\times1)$
surface measured at room temperature. Numbers highlight visible FAU
atoms. The white rhomb and down triangle mark the weak spots originating
from the buckled dimers in the trench. The structure of the pentamer
with an interstitial atom is shown by black lines and balls for eye
guidance. (a) $U=-0.8\,\mathrm{V}$ (filled electronic states). (b)
$U=+0.8\,\mathrm{V}$ (empty electronic states). Reproduced with permission
from Ref.~\onlinecite{tey17}. Copyright 2017 Pleiades Publishing,
Ltd.}
\end{figure}

\begin{figure}
\includegraphics[width=6.5cm]{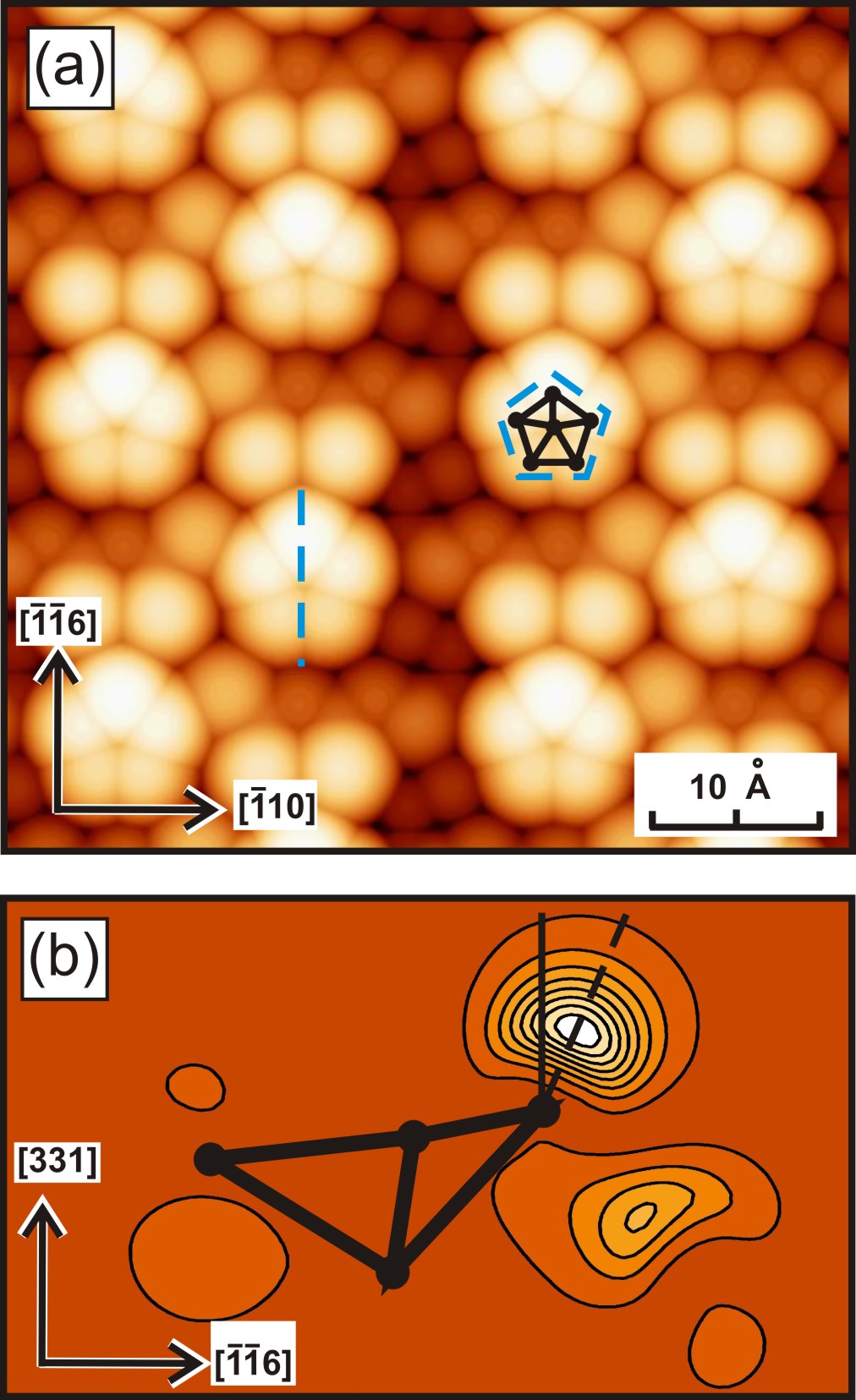}

\caption{\label{fig3} (a) Calculated STM image of the Si$(331)$ surface assuming
the revised 8P model for the $(12\times1)$ reconstruction proposed
in the present study. The structure of the pentamer with an interstitial
atom is shown by black lines and balls for eye guidance. The pentagon
shown by a dashed blue line illustrates the apparent size of the pentamer
feature, as derived from the center of bright spots. The standard
bulk-optimized DZP basis set is used for the two upper bilayers of
the slab (cut-off radii for \emph{s-}, \emph{p-}, and \emph{d-}orbitals
are: $R_{s}=5.7\,\mathrm{Bohr}$, $R_{pd}=7.3\,\mathrm{Bohr}$). The
bias voltage corresponds to $+0.8\,\mathrm{eV}$ with respect to the
theoretical Fermi level (empty electronic states). The calculated
STM image for $U=-0.8$ V bias (filled electronic states) is very
similar to that shown in this figure. (b) The vertical cut of calculated
LDOS isosurfaces along the blue dashed line crossing pentamer in (a),
integrated over a 0.8 eV energy window above the calculated Fermi
level. The projection of a pentamer atomic structure is overlaid.
The solid black line shows a vertical direction with respect to the
surface. The dashed line is drawn from the apex atom along the direction
of maximum intensity of the nearest bright spot. }
\end{figure}

Figure~\ref{fig3}(a) shows the calculated constant-current STM image
of the Si$(331)\textrm{-}(12\times1)$ according to the revised 8P
model of Fig.~\ref{fig1}(c). The standard bulk-optimized DZP basis
set was used for Si surface atoms. The structure of the pentamer with
an interstitial atom is also overlaid in Fig.~\ref{fig3}(a). The
actual distance between the nearest Si atoms in the pentamer is $2.3\,\mathrm{\mathring{A}}$.
The dashed blue pentamer in Fig.~\ref{fig3}(a) emphasizes its apparent
size as derived from the calculated STM image. The distance between
two nearest spots is about $3.5\,\mathrm{\mathring{A}}$, in agreement
with the analogous distance as derived from the experimental STM image
of Fig.~\ref{fig2}(b). On the other hand this figure corresponds
to a deviation of more than 50\% with respect to the atomic positions
according to the 8P model. Such a size mismatch sustained an argument
against the atomic models of $(331)$ silicon surface involving five-fold
rings of Si atoms.\citep{tey17} It should be noted, however, that
STM is a technique sensitive to LDOS, rather than the positions of
atomic nuclei.\citep{ter85}

Figure~\ref{fig3}(b) shows a vertical cut of the integrated LDOS
(between +0.8 V and Fermi level) along the blue dashed segment crossing
the pentamer in Fig.~\ref{fig3}(a), combined with a projection of
a pentamer atomic structure. The brightest spot at the upper right
corner of Fig.~\ref{fig3}(b) shows a high intensity for the empty
LDOS located near the Si radical at the upmost apex of the pentamer
in Fig.~\ref{fig3}(a). Clearly, the radical state does not point
upwards (along $[331]$), rather making an angle of about 22° with
respect to the surface normal and away from the center of the pentamer.
Since STM is intrinsically sensitive to LDOS\citep{ter85} and because
the tip usually scans between 4 to 10 Å above the surface,\citep{hof03}
it becomes evident that the slanted radicals will project a magnified
image of the underlying pentamer. This explains the apparent contradiction
between the size of the pentamers from the atomistic model and those
derived directly from STM images (see Ref.~\onlinecite{tey17}).

The main advantage of using atomic orbitals as a basis set resides
on their efficiency (fewer orbitals per state are needed in order
to obtain a precision similar to that of plane-wave codes). However,
there are also known issues when atomic orbitals are used for the
calculation of the surface properties, such as surface formation energies,
surface state energies and work functions.\citep{san09} In this paper,
we show that, in addition to the problems listed above, the calculated
STM images may show no bias dependency because of the short decay
length of the basis functions into the vacuum space. However, we demonstrate
that an accurate description of the STM images is, nevertheless, possible
with an appropriately tuned basis set. 

First, we note that although spot positions in the calculated STM
image of Fig.~\ref{fig3}(a) closely reproduce the spot positions
in the experimental STM images (Figs.~\ref{fig2}(a) and \ref{fig2}(b)),
their relative intensities are incorrect. Second, the positive and
negative bias STM images calculated using the standard bulk-optimized
DZP basis set are almost undistinguishable, while the experimental
STM images show a strong bias-dependency (Figs.~\ref{fig2}(a) and
\ref{fig2}(b)).\citep{tey17} The height contrast measured from the
calculated STM image shown in Fig.~\ref{fig3}(a) mostly represents
the surface topography. It lacks the electronic contribution, closely
matching the height differences measured from Si nuclei positions.

The Sankey-Niklewski-type orbitals used in the \textsc{siesta} code
are the eigenfunctions of the (pseudo) atom within a sphere of cut-off
radius \emph{R} (orbitals are strictly zero beyond this radius).\citep{sol02}
Thus, the orbitals are obtained as a product between a confined radial
function and a spherical harmonic. In order to describe the properties
of the surfaces accurately, one needs to include orbitals which can
describe the long decay of the wave functions.

Three different schemes were proposed in Ref.~\onlinecite{san09}.
First is to add diffuse functions (with a longer cut-off than that
of the bulk-optimized orbitals) to the basis. According to our test
calculations, this scheme leads to improved calculated STM images.
However, the overlap of the extra orbitals with the orbitals of the
original bulk-optimized basis set leads to artifacts in calculated
images. Another scheme is to add a set of floating orbitals (ghost
atoms) located above the surface. These are off-site orbitals centered
at the points where there are no atoms. The problem with this scheme
is that, on the reconstructed surfaces, such as the one shown in Fig.~\ref{fig1}(e),
there is no obvious and unbiased way for their placement. The resulting
STM images also show artifacts which additionally depend on the placement
of off-site orbitals.

\begin{figure}
\includegraphics[width=6.5cm]{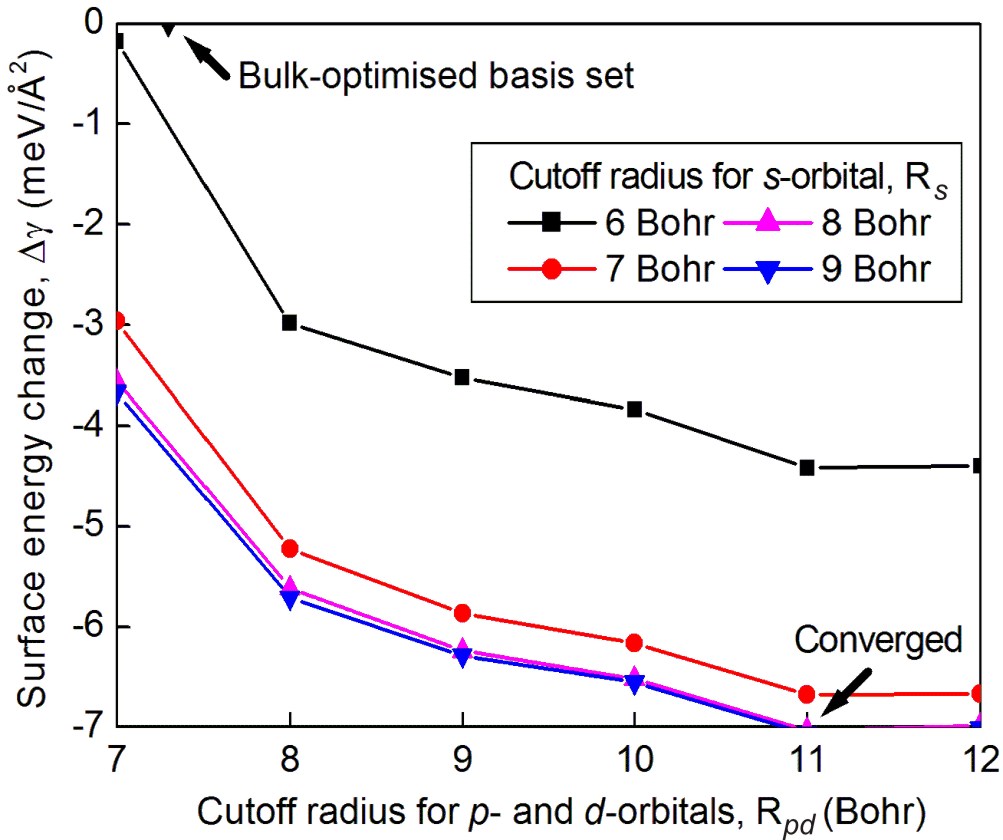}

\caption{\label{fig4} The Si$(331)\textrm{-}(12\times1)$ surface energy change
as a function of the cut-off radii for \emph{s-}, \emph{p-}, and \emph{d-}orbitals
in the two upper bilayers of the slab.}
\end{figure}

The last scheme is to enlarge the cut-off radius of the atomic orbitals
at the surface layer atoms only, which allows the wave functions to
spread further into vacuum. Increasing the orbital cut-off radii at
the surface atoms leads to the basis functions with the natural decay
of the atomic orbitals in the free atom which greatly improves the
description of the surface properties.\citep{san09} The confinement
radii for the atoms in the two upper bilayers of the slab were taken
as a variational parameter to optimize the quality of the basis set
and are chosen to minimize the slab total energy. The radii for \emph{s-}
and \emph{p-}orbitals ($R_{s}$ and $R_{p}$) were varied independently,
while the radius for the \emph{d-}orbital was set equal to that of
the \emph{p-}orbital ($R_{p}=R_{d}=R_{pd}$). The change of the slab
total energy is represented as corresponding change of the relative
surface energy in Fig.~\ref{fig4}. According to that figure, the
total energy of the slab shows a noticeable reduction when the confinement
radii are increased, and that clearly indicates the need to expand
the basis to describe properly the vacuum region. The energy converges
at about $R_{s}=9\,\mathrm{Bohr}$ and $R_{pd}=11\,\mathrm{Bohr}$
(Fig.~\ref{fig4}). The bias-dependent STM images in Figs.~\ref{fig5}(a)
and \ref{fig5}(b) are calculated using the DZP basis set optimized
for the surface according to the described procedure. These images
closely reproduce the bias dependence of the experimental STM images
in Figs.~\ref{fig2}(a) and \ref{fig2}(b), and that strongly supports
the revised 8P model for the Si$(331)\textrm{-}(12\times1)$ surface.
The STM images for the lowest energy configuration were also calculated
using the \textsc{vasp} code and within the GGA to the exchange-correlation
interactions, essentially resulting in identical STM images as those
in Figs.~\ref{fig5}(a) and \ref{fig5}(b) (not shown).

It is interesting to compare the absolute surface energy calculated
using the standard bulk-optimized DZP basis with that calculated using
the surface-optimized DZP basis, as well as with a well-converged
PW calculation. The absolute surface energy of the unreconstructed
and unrelaxed Si$(331)$ surface is $127.2\,\mathrm{meV/\mathring{A}^{2}}$
when using the PW code.\citep{bat09a} We used a 14 bilayers thick
symmetric slab to calculate the formation energy of the same surface
using a strictly localized DZP basis set. We found that the formation
energy of that surface is $129.8\,\mathrm{meV/\mathring{A}^{2}}$
when using the standard bulk-optimized basis set, while it is $126.6\,\mathrm{meV/\mathring{A}^{2}}$
when using the surface-optimized basis. Thus, besides improving significantly
the quality of the calculated STM images, the surface-optimized basis
improves the absolute surface formation energy, leaving the relative
surface energies essentially unchanged (see Tab.~\ref{tab1}).

\subsection{Origin of STM contrasts and electronic structure of the Si(331)-(12\texttimes 1)
surface\label{subsec:contrasts}}

\begin{figure}
\includegraphics[width=6.5cm]{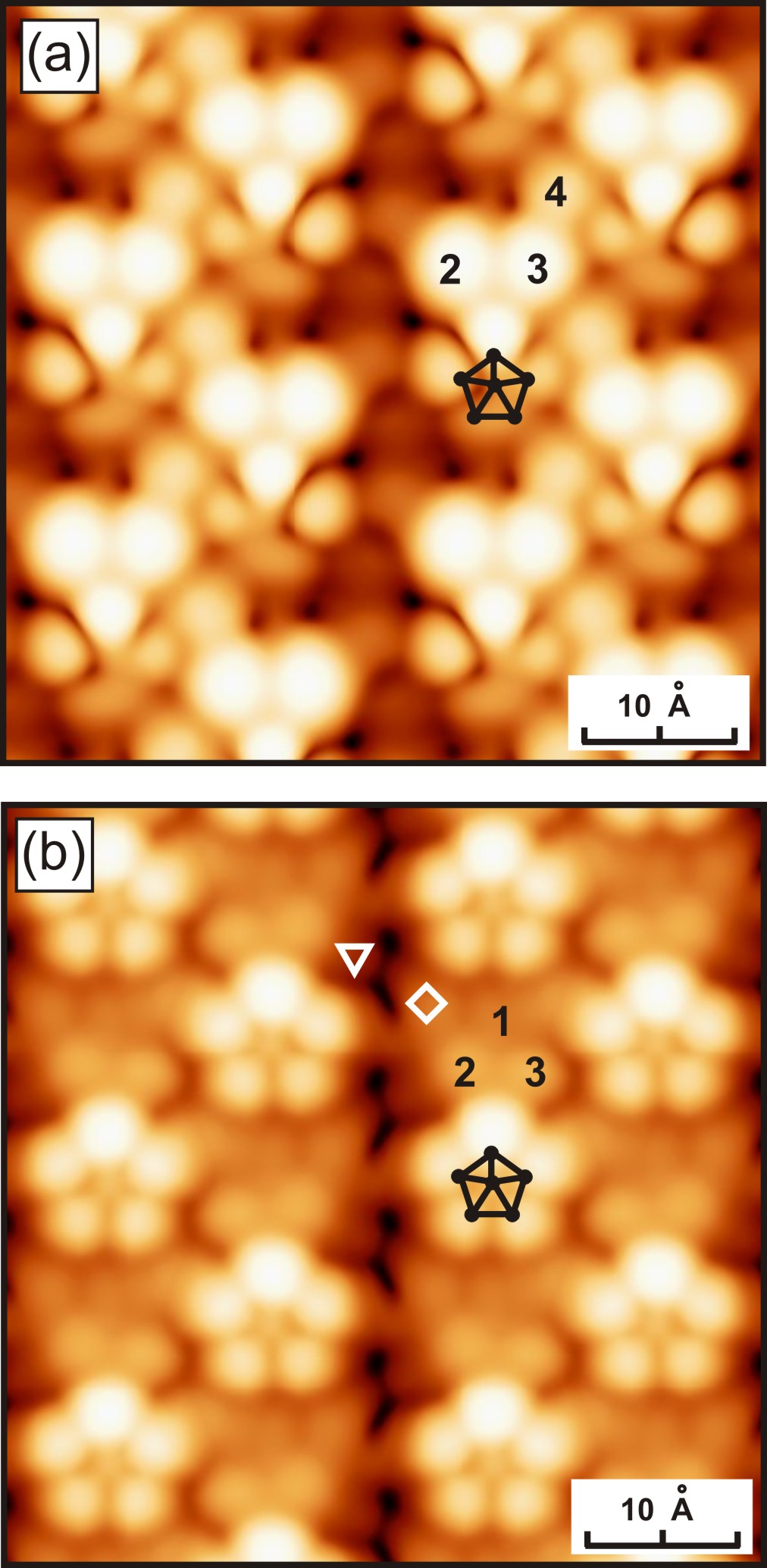}

\caption{\label{fig5} Calculated STM images of the Si$(331)$ surface assuming
the revised 8P model for the $(12\times1)$ reconstruction. A surface-optimized
DZP basis set was used for two upper bilayers of the slab (cut-off
radii for \emph{s-}, \emph{p-}, and \emph{d-}orbitals are: $R_{s}=9\,\mathrm{Bohr}$,
$R_{pd}=11\,\mathrm{Bohr}$). The numbers highlight visible FAU atoms.
The white rhomb and down triangle mark the weak spots originating
from the buckled dimers in the trench. The structure of a pentamer
with an interstitial atom is shown as black lines and balls for eye
guidance. (a) The bias voltage corresponds to $-0.8\,\mathrm{eV}$
with respect to the theoretical Fermi level (filled electronic states).
(b) The bias voltage corresponds to $+0.8\,\mathrm{eV}$ with respect
to the theoretical Fermi level (empty electronic states).}
\end{figure}

In figures~\ref{fig5}(a) and \ref{fig5}(b) we depict the calculated
STM images of the Si$(331)$ surface assuming the revised 8P model
for the $(12\times1)$ reconstruction. The surface-optimized basis
set is used for two upper bilayers of the slab and the lowest energy
buckled configuration shown in Fig.~\ref{fig1}(e) is assumed. The
images were calculated using a tip-surface separation of about $4.8$~Å
(as measured from the topmost point of the LDOS isosurface to the
highest point of the Si$(331)$ lattice). We found that the dependence
of the STM images on the tip-surface separation is weak, with spots
becoming more diffuse as the distance increased within the range $3.0\textrm{-}5.6$~Å
(setting different LDOS isosurfaces). The positive and negative bias
STM images differ from each other in relative intensity of spots.
This is in agreement with the experimental STM images shown in Figs.~\ref{fig2}(a)
and \ref{fig2}(b) (Ref.~\onlinecite{tey17}), providing compelling
evidence of a strong contribution of the electronic states to the
observed STM height contrast.

There is a clear similarity between the experimental and calculated
STM images shown in Figs.~\ref{fig2}(a), (b) and Figs.~\ref{fig5}(a),
(b), respectively. The calculated STM image of the pentamer with an
interstitial atom shows 5 bright spots at positive bias (Fig.~\ref{fig5}(b))
and 4 spots at negative bias (Fig.~\ref{fig5}(a)) in agreement with
the bias-dependent experimental STM images of the pentamers on the
Si$(110)$ surface.\citep{an00} The experimental STM image of the
pentamer on the Si$(331)$ surface shows 5 distinct spots at the positive
bias (Fig.~\ref{fig2}(b)), while only 3 spots are visible in the
negative bias (Fig.~\ref{fig2}(a)). We should note here that our
calculated positive bias STM images are closer to the images reported
by Battaglia \emph{et al.} in Refs.~\onlinecite{bat09,bat09a,bat11}
than to the images reported by Teys in Ref.~\onlinecite{tey17}.
The difference between experimental STM images by different authors
can, in principle, be explained by variations in the quality of the
STM tips, which is hard to control in experiments. We, therefore,
suggest that the small difference observed between the experimental
and calculated STM images in Figs.~\ref{fig2}(a) and Figs.~\ref{fig5}(a)
could be due to the STM tip. In other words, the 4-th banana-shaped
spot observed in the calculated STM image of the pentamer in Fig.~\ref{fig5}(a),
is simply unresolved in the experimental STM image of Fig.~\ref{fig2}(a).

Depending on the applied bias polarity, different spots are highlighted
in the STM images of FAU (Figs.~\ref{fig2} and \ref{fig5}). These
spots are numbered according to the scheme shown in Fig.~\ref{fig1}(e).
In the STM images with a negative bias (filled electronic states),
spots 2, 3 and 4 are bright, while spot 1 is dimmed (Figs.~\ref{fig2}(a)
and \ref{fig5}(a)). In the STM images with a positive bias (empty
electronic states), spots 1, 2 and 3 can be distinguished, while spot
4 is less visible (Figs.~\ref{fig2}(b) and \ref{fig5}(b)). Thus,
the relative change of the spot brightness in the experimental STM
images of FAUs upon reversing the bias polarity is reproduced in the
theoretical STM images.

Few weak spots are visible in the trench observed as a dark vertical
stripe in the middle of the experimental STM image of Fig.~\ref{fig2}(b).
These spots are marked by a white rhomb and triangle. According to
the revised 8P model shown in Fig.~\ref{fig1}(e), these spots originate
from the buckled dimers located in the trenches between the pentamers.
The spots are well reproduced in the calculated STM image assuming
the revised 8P model for the Si$(331)\textrm{-}(12\times1)$ surface
in Fig.~\ref{fig5}(b). The remaining small differences between the
experimental and calculated STM images might be explained by the dynamic
buckling of dimers and FAUs and, therefore, by considering the contributions
from other (higher energy) buckled configurations listed in Tab~\ref{tab1}.

\begin{figure}
\includegraphics[width=6.5cm]{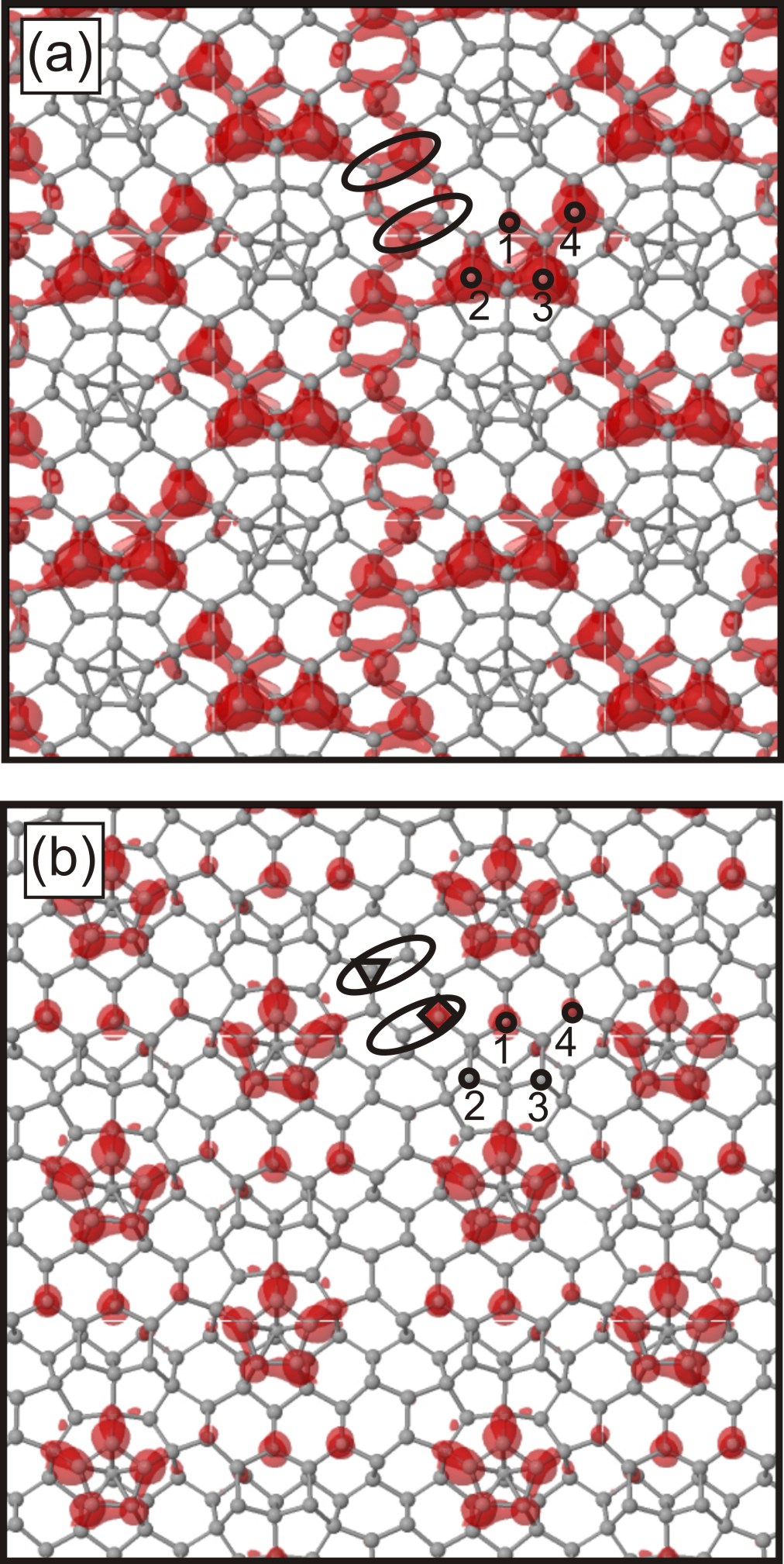}

\caption{\label{fig6}LDOS isosurfaces integrated in a $0.8\,\mathrm{eV}$
energy window below (a) and above (b) the calculated Fermi level.
The isosurface value is $7\times10^{-4}\,\mathrm{\mathring{A}^{-3}}$
for both images. Dimers are highlighted by ovals, FAU atoms are numbered
1 - 4. These results were obtained within the hybrid DFT. The rhomb
and triangle mark the surface atoms in the trench related to the weak
spots visible in the STM images of Figs.~\ref{fig2}(b) and \ref{fig5}(b).
The revised atomic 8P model for the Si$(331)\textrm{-}(12\times1)$
surface is overlaid.}
\end{figure}

According to the Tersoff-Hamann approximation, the STM images represent
a combination of surface topography and a map of the sample surface
LDOS.\citep{ter85,hof03} It is therefore interesting to plot LDOS
isosurfaces in order to isolate the electronic contribution to the
STM images. Figure~\ref{fig6} shows the calculated LDOS isosurfaces
integrated over a $0.8\,\mathrm{eV}$ window just below (a) and above
(b) the calculated Fermi level (filled and empty electronic states,
respectively). The lowest energy buckled configuration shown in Fig.~\ref{fig1}(e)
is assumed. These results were obtained within hybrid DFT (using the
HSE06 functional), allowing a direct comparison with the DOS data
of Figure~\ref{fig7} below. One can see that the filled and empty
LDOS images are fully complementary.

The empty LDOS are located on the pentamer atoms and 3-fold coordinated
atoms having the \emph{sp}$^{2}$-like (planar) configuration of their
bonds. These latter atoms correspond to the depressed atoms in the
buckled configuration shown in Fig.~\ref{fig1}(e). The filled LDOS
are located on the 3-fold coordinated atoms showing \emph{sp}$^{3}$-like
configurations of their bonds (both in FAUs and on the protruded dimer
atoms in the trench). These atoms correspond to the protruded atoms
in the buckled configuration shown in Fig.~\ref{fig1}(e). The overall
redistribution of electron density from the \emph{sp}$^{2}$-like
to \emph{sp}$^{3}$-like Si surface atoms is in agreement with the
observations made on $(100)$ and $(7\,7\,10)$ silicon surfaces.\citep{bech03,zha14}

Finally, we studied the band structure of the Si$(331)$ reconstructed
surface using hybrid DFT. This allows us to make a direct comparison
with the spectroscopic data involving the energies in the range of
the band gap. Again, as in the case of STM images, we consider only
the lowest energy buckled configuration shown in Fig.~\ref{fig1}(e)
(first configuration in Tab.~\ref{tab1}). Due to the sparse nature
of the electronic states within the gap (as evidenced by Fig.~\ref{fig6}),
the bands show very weak dispersion (not shown here). The calculated
total DOS is plotted in Fig.~\ref{fig7}. The spectrum shows two
broad peaks, when using a $0.15\,\mathrm{eV}$ broadening (thick red
line): one above the valence band maximum, and another one below the
conduction band minimum. The calculated band gap for the surface is
$0.58$~eV, which is almost twice the value obtained using LDA (Table~\ref{tab1}).
The shape of the 0.15~eV broadened DOS spectrum and the calculated
value of the band gap are in a very good agreement with the experimental
analogues from angle-resolved photoelectron and scanning tunneling
spectroscopy (STS). These also show two broad peaks edging the band
gap, whose width was estimated between $0.5\,\mathrm{eV}$ and $0.6\,\mathrm{eV}$.\citep{bat11}

Concerning the broadening of the DOS peaks, we note that the error
of the DOS should be close to that of the calculated gap within hybrid
DFT as reported in Section II, which is about $0.03$~eV. The thermal
smearing due to finite temperature (at room temperature) is about
$0.03$~eV. Both these values are lower than the $0.15$~eV peak
smearing of Fig.~\ref{fig7}, therefore accounting for the broadening
due to experimental conditions. The typical STS resolution is about
$0.2$~eV, while some disorder on the Si$(331)$ surface and dynamic
buckling will contribute to a smeared spectrum (\emph{c.f.} reported
LDA band gaps in Tab.~\ref{tab1}). The broad peaks calculated at
$0.15$~eV smearing hide a fine structure when the spectrum is calculated
using a $0.10$~eV broadening (thin black line). This finer spectrum
is given for guidance, hoping for a future improvement in equipment
resolution and sample quality.

\begin{figure}
\includegraphics[width=6.5cm]{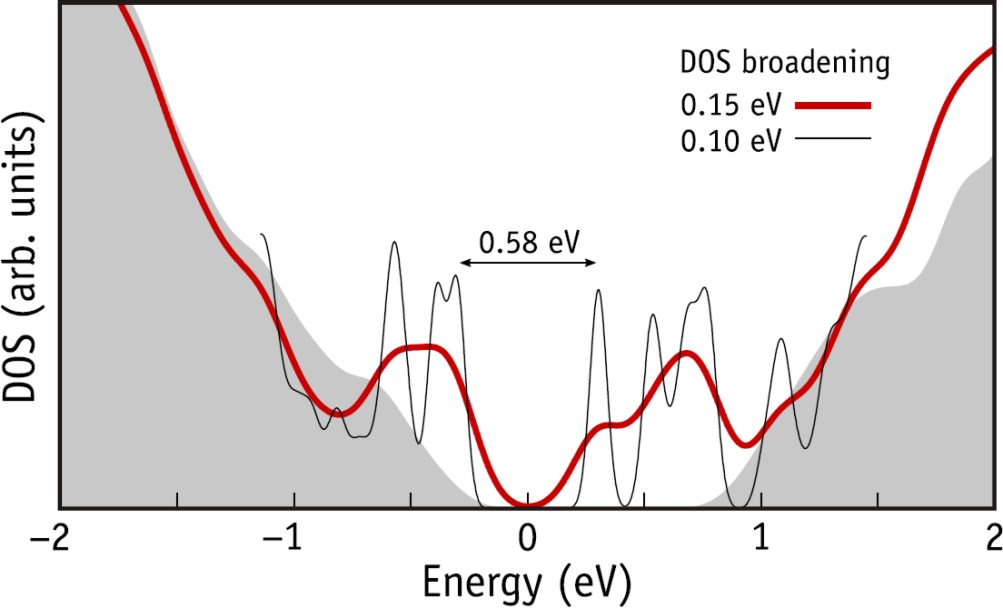}

\caption{\label{fig7}Total DOS plot of the Si$(331)-(12\times1)$ slab. The
gray shaded area shows the calculated DOS of the silicon bulk. These
results were obtained within the hybrid DFT.}
\end{figure}

\section{Conclusions}

In summary, the Si$(331)\textrm{-}(12\times1)$ surface reconstruction
was studied theoretically using density functional calculations, and
the results were compared to the available experimental data. We propose
important correction to the 8P structural model of the Si$(331)$
reconstructed surface, suggesting that the trenches between the zig-zag
chains of 8PUs consist of dimers. It is demonstrated that an accurate
description of the bias-dependent STM images and energetics of the
reconstructed surface is possible using a basis set consisting of
atomic orbitals. The apparent contradiction between the size of the
pentamers from the atomistic model and those derived directly from
STM images is explained within the context of the Tersoff-Hamann model.
The revised 8P model of the Si$(331)$ reconstructed surface shows
a remarkable agreement with the experimental STM images, spectroscopy
data and the width of the surface band gap. The calculated energy
barriers separating different buckled configurations of the Si$(331)$
surface assume that this surface is prone to dynamic buckling at room
temperature. Overall, the presented results demonstrate that the revised
8P model of the Si$(331)$ reconstructed surface is in a very good
agreement with the available experimental data and, therefore, it
can serve as a reliable starting point for future research studies
related to this surface.
\begin{acknowledgments}
We would like to thank the Novosibirsk State University for providing
the computational resources. R. Zhachuk acknowledges the financial
support of the Russian Foundation for Basic Research (Project No.
18-02-00025). This work was partially funded by the Fundação para
a Ciência e a Tecnologia (FCT) under the contract UID/CTM/50025/2013
and by FEDER funds through the COMPETE 2020 Program. Partial funding
was also provided by the Slovak Academy of Sciences SASPRO program
1239/02/01, and by the National Research Development and Innovation
Office of Hungary project FK124100.
\end{acknowledgments}

\section*{APPENDIX: Calculated STM images and DOS of the original 8P model}

\begin{figure}
\includegraphics[width=6.5cm]{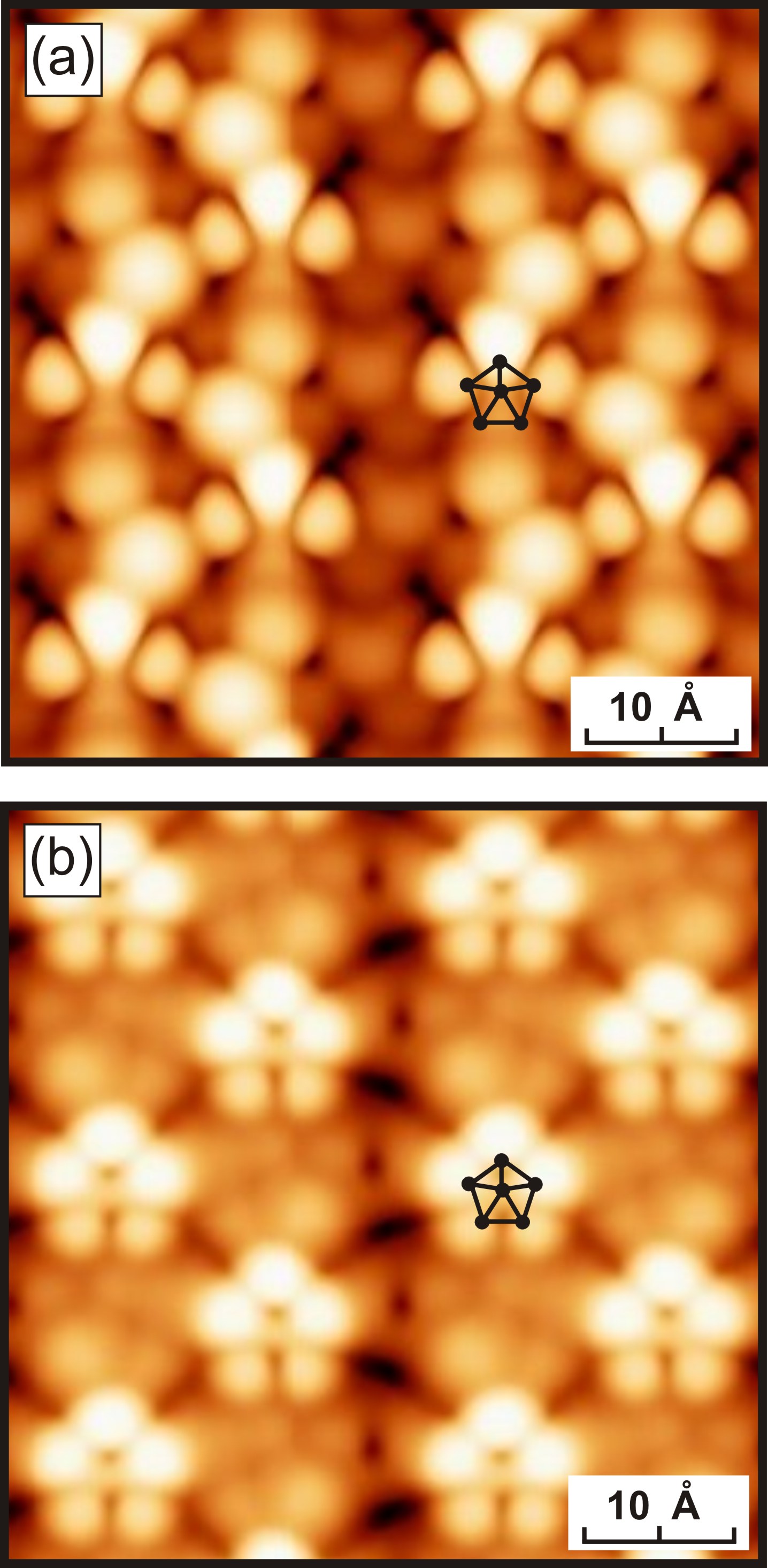}

\caption{\label{fig8}Calculated STM images of the Si$(331)$ surface assuming
the original 8P model for the $(12\times1)$ reconstruction, proposed
in Ref.~\onlinecite{zha17}. Surface-optimized DZP basis set is used
for two upper bilayers of the slab (cut-off radii for \emph{s-}, \emph{p-},
and \emph{d-}orbitals are: $R_{s}=9\,\mathrm{Bohr}$, $R_{pd}=11\,\mathrm{Bohr}$).
The structure of a pentamer with an interstitial atom is shown as
black lines and balls for eye guidance. (a) The bias voltage corresponds
to $-0.8\,\mathrm{eV}$ with respect to the theoretical Fermi level
(filled electronic states). (b) The bias voltage corresponds to $+0.8\,\mathrm{eV}$
with respect to the theoretical Fermi level (empty electronic states).}
\end{figure}

\begin{figure}
\includegraphics[width=6.5cm]{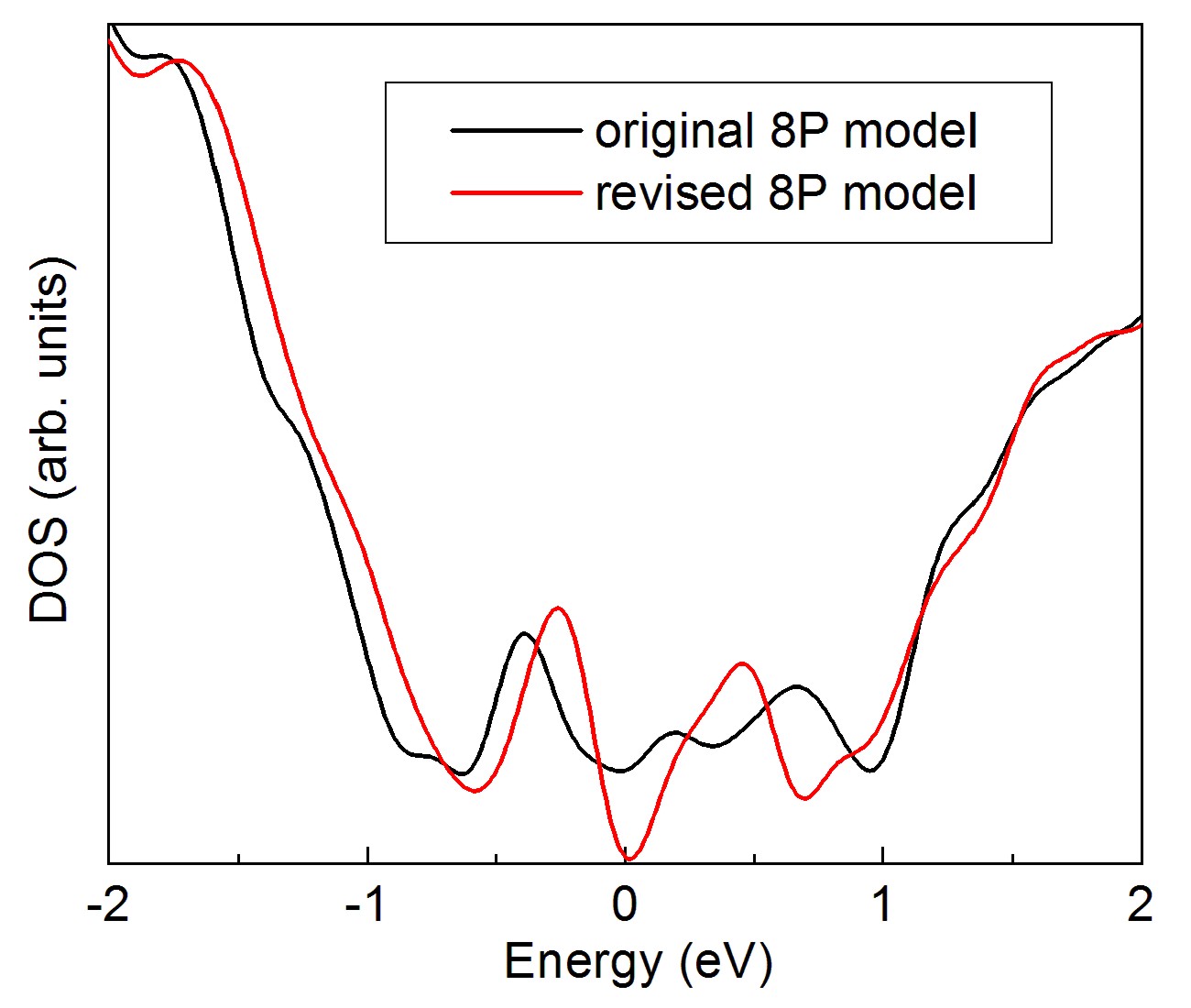}

\caption{\label{fig9}Total DOS plot of the Si$(331)\textrm{-}(12\times1)$
slab according to the original 8P model from Ref.~\onlinecite{zha17}
(black line) and according to the revised 8P model (red line), proposed
in the present work. A $0.15\,\mathrm{eV}$ peak broadening was used
for both spectra.}
\end{figure}

In this section we show calculated STM images (Fig.~\ref{fig8})
and DOS (Fig.~\ref{fig9}) of the original 8P model for the Si$(331)\textrm{-}(12\times1)$
surface, as proposed in Ref.~\onlinecite{zha17}. Figure~\ref{fig9}
compares the DOS for the original and the newly proposed 8P models.
Both calculations were carried out using the \textsc{siesta} code
within the LDA to the exchange-correlation energy.

Figures \ref{fig8}(a) and \ref{fig8}(b) show the calculated STM
images of the Si$(331)$ surface assuming the original 8P model for
the $(12\times1)$ reconstruction, proposed in Ref.~\onlinecite{zha17}.
The surface-optimized DZP basis set used for the revised model shown
in Fig.~\ref{fig5} is used in this case as well. The calculated
STM image in Fig.~\ref{fig8}(a) shows that the relative intensity
of spots in FAUs does not match the relative intensity of spots in
corresponding experimental STM image in Fig.~\ref{fig2}(a). More
importantly, the spot marked by a down triangle in the experimental
STM image in Fig.~\ref{fig2}(b) is completely absent in the calculated
STM image of Fig.~\ref{fig8}(b).

Figure~\ref{fig9} shows the calculated DOS spectra according to
the original (black line) and revised (red line) 8P models of the
Si$(331)\textrm{-}(12\times1)$ surface, respectively. The DOS spectrum
calculated using the original 8P model shows a non-vanishing DOS amplitude
in the region of the Fermi energy ($E=0$~eV), at variance with the
experimental data.\citep{bat11}

Hence, Figs.~\ref{fig8} and \ref{fig9} demonstrate that the original
8P model for the Si$(331)\textrm{-}(12\times1)$ surface cannot account
for the observations, unlike the revised 8P model.

\bibliographystyle{apsrev4-1}
%\bibliography{refs}


\end{document}